\documentclass[runningheads]{llncs}
\usepackage[T1]{fontenc}
\usepackage{multirow}
\usepackage{amsmath}
\usepackage{enumitem}   
\usepackage{graphicx}
\usepackage{bbding}
\begin{document}
\title{Are Transformers in Pre-trained LM A Good ASR Encoder? An Empirical Study}
\titlerunning{PLM as ASR encoder}
\author{Keyu An, Shiliang Zhang, Zhijie Yan}
\authorrunning{Keyu An et al.}
\institute{Speech Lab, Alibaba Group, China \\
\email{{ankeyu.aky,sly.zsl,zhijie.yzj}@alibaba-inc.com}}
\maketitle              % typeset the header of the contribution
\begin{abstract}
In this study, we delve into the efficacy of transformers within pre-trained language models (PLMs) when repurposed as encoders for Automatic Speech Recognition (ASR). Our underlying hypothesis posits that, despite being initially trained on text-based corpora, these transformers possess a remarkable capacity to extract effective features from the input sequence. This inherent capability, we argue, is transferrable to speech data, thereby augmenting the acoustic modeling ability of ASR. Through rigorous empirical analysis, our findings reveal a notable improvement in Character Error Rate (CER) and Word Error Rate (WER) across diverse ASR tasks when transformers from pre-trained LMs are incorporated. Particularly, they serve as an advantageous starting point for initializing ASR encoders. Furthermore, we uncover that these transformers, when integrated into a well-established ASR encoder, can significantly boost performance, especially in scenarios where profound semantic comprehension is pivotal. This underscores the potential of leveraging the semantic prowess embedded within pre-trained transformers to advance ASR systems' capabilities.

\end{abstract}
\section{Introduction}
Textual pre-trained language models (PLMs) have attained unprecedented achievements across many natural language processing (NLP) tasks~\cite{gpt4,qwen}, sparking an interest in harnessing their generative prowess to augment Automatic Speech Recognition (ASR) systems. This recent development manifests in two principal strategies: The \textbf{post-processing} methodology employs PLMs to refine ASR outputs, either through language model fusion in decoding~\cite{fusion} or by deploying them for error rectification post-ASR decoding~\cite{asr_correct}. The \textbf{multi-modal} paradigm introduces innovative architectural designs that synergize an acoustic signal encoder with a PLM-based decoder, typically in a decoder-only configuration. Noteworthy cross-modal connectors include using Q-former~\cite{salmonn} or a simple linear layer~\cite{qwen_audio}. These mechanisms empower PLMs to transcend the text-only modality, enabling them to process and understand speech data directly.

Diverging from conventional methodologies, we introduce a novel strategy to utilize the capabilities of pre-trained language models (PLMs) for the advancement of Automatic Speech Recognition (ASR). Acknowledging the proficiency of transformer~\cite{self-attention} layers within PLMs in modeling relationships among textual elements, we posit that this capability can generalize to acoustic signals, thereby benefiting acoustic modeling. The central objective of this paper is to scrutinize the viability of employing transformers extracted from pre-trained LMs as dedicated ASR encoders. To this end, we pose two pivotal inquiries: (1) Can the transformers from pre-trained LMs, coupled with a rudimentary convolutional input layer, constitute an efficacious ASR encoder? and (2) To what extent can these transformers augment the performance of an already well-trained ASR encoder, especially regarding its semantic modeling capabilities? Our empirical exploration, conducted on datasets including AISHELL-1, Librispeech, and an in-house dataset, illuminates several advantageous characteristics of PLM-derived transformers for ASR. These findings not only validate the aforementioned hypotheses but also illuminate new methods for integrating PLM into ASR systems.

\section{Transformer in pre-trained LM as ASR encoder}
\label{method}
\begin{figure}
\centering
\includegraphics[width=0.5\textwidth]{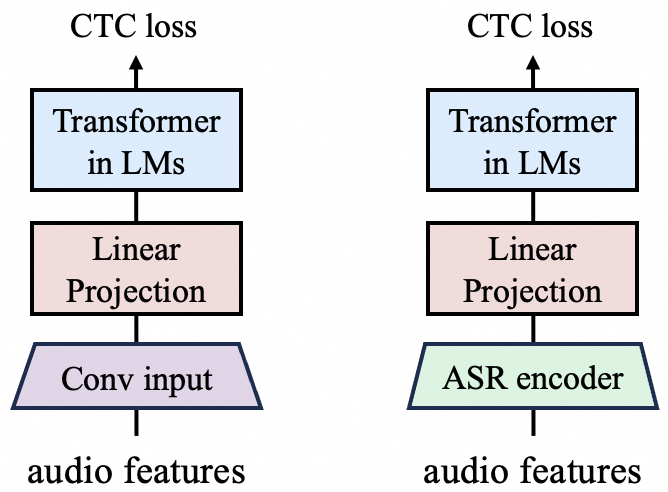}
\caption{Transformer in pre-trained LM as ASR encoder. } \label{fig1}
\end{figure}
Our proposed ASR model adopts an encoder-only architecture, based on Connectionist Temporal Classification (CTC)~\cite{ctc}, with the Qwen model's~\cite{qwen} transformers serving as the backbone of the encoder. This encoder implementation takes two forms: either by initiating with a convolutional input layer followed by Qwen transformers, or by integrating Qwen transformers on top of an existing, pre-trained ASR encoder.

Formally, given an input audio feature sequence ${\mathbf X}$ and the corresponding target label sequence ${\mathbf y}$, the objective function defined by the CTC loss is:

\begin{equation}
  \mathcal{L} = - {\rm log}P_{\rm CTC}({\mathbf y}|{\rm softmax} ({\rm Linear_{CTC}}( {\rm Enc}({\mathbf X}))))  
\end{equation}

Here, ${\rm Enc}({\mathbf X})$ represents the encoding process, which is realized in one of two configurations:

\begin{equation}
{\rm Enc}({\mathbf X}) = {\rm Qwen}({\rm Linear}({\rm Conv}({\mathbf X}))) 
\end{equation}
or 
\begin{equation}
{\rm Enc}({\mathbf X}) ={\rm Qwen}({\rm Linear}({\rm Enc_{ASR}}({\mathbf X}))) 
\end{equation}

In these formulations, ${\rm Qwen}$ signifies the transformer layers inherent to the Qwen model, designed originally for text generation but adapted here for acoustic feature processing. Notably, the causal attention masks typically employed for autoregressive decoders are dispensed with, as they are unnecessary for the encoder's operation in our ASR framework. A linear transformation layer is employed to facilitate the transition between the convolutional front-end or the pre-trained  ASR encoder and the Qwen transformers.

\section{Experiments}
\subsection{Experiment settings}
To thoroughly evaluate the effectiveness of our approach, we carry out experiments across three diverse datasets: the publicly accessible 170-hour Mandarin speech dataset AISHELL-1 and 960-hour English corpus LibriSpeech, and an in-house Mandarin dataset encompassing 60,000 hours of speech data. Uniformly across these datasets, we adopt filterbanks as the model's input features, computed over 25ms windows with a 10ms frame shift.

In the variant of our model incorporating a convolutional input layer, the input sequences undergo a 4-fold downsampling via convolutional strides. For the model that leverages a pre-trained ASR encoder, a stacking of consecutive frames is performed before applying a downsampling rate of 6. This pre-trained ASR encoder is a 50-layer Memory Equipped Self-Attention Network (SAN-M)~\cite{sanm}, which has been previously trained on our large-scale 60,000-hour Mandarin dataset. Unless specified otherwise, we employ Qwen-1.5-0.5B~\footnote{https://huggingface.co/Qwen/Qwen1.5-0.5B} as the foundation model. In integrating Qwen, we omit its embedding and output layers, focusing exclusively on the transformer layers. All configurations of our models are coupled with a greedy CTC decoder, aligning with the encoder-only architecture. This design choice underscores our commitment to exploring the potential of transformers in PLMs for enhanced ASR encoders.
\subsection{Results}
To systematically assess the capacity of transformers derived from pre-trained language models (PLMs) in acoustic modeling, we establish a comparative evaluation across several model configurations. These configurations include: 

(1) Model with Convolutional Input Only: Serves as a baseline, utilizing solely a convolutional layer for feature extraction and encoding. 

(2) Frozen Randomly Initialized Transformers: Features from the convolutional input layer are fed into transformers architecturally akin to those in Qwen, but with randomly initialized weights. These transformer layers remain static (frozen) during training. 

(3) Frozen Pre-Trained Transformers: Similar setup to the previous model, except the transformers now incorporate pre-trained weights from Qwen. Despite this, these pre-trained layers are also frozen during ASR training. 

(4) Jointly Optimized Random Transformers: Here, transformers with random initializations are stacked atop the convolutional layer. Unlike before, both the transformers and the convolutional layer are fine-tuned jointly based on the ASR loss. 

(5) Jointly Optimized Pre-Trained Transformers: This configuration mirrors the previous one but employs transformers initialized with pre-trained parameters from Qwen. Both the Qwen transformers and the convolutional input layer are fine-tuned jointly using the ASR loss. 

Performance metrics for these models are summarized in Tables~\ref{tab1} and \ref{tab2}. The outcomes unambiguously highlight several key observations:
\begin{enumerate}[label=(\roman*)]
\item \textbf{Insufficiency of Convolutional Layer Alone}: Relying solely on a convolutional input layer for acoustic modeling is inadequate, highlighting the need for more advanced structures to effectively capture the intricacies of speech data.
\item \textbf{Potential of Random Initialized Transformers}: The addition of transformer layers, even when randomly initialized and without ASR-specific optimization, leads to substantial improvement. This suggests that the inherent architecture of transformers, even without specialized training, possesses a foundational capability to model relationships between inputs, contributing positively to acoustic modeling.
\item \textbf{Superiority of Pre-Training}: Models using pre-trained Qwen significantly outperform those initialized randomly, regardless of whether the Qwen parameters are frozen or not. This emphasizes that transformers within pre-trained LMs, despite being trained on text alone, are adept at capturing contextual information in inputs, and this ability generalizes to speech inputs. Hence, pre-trained LMs offer a robust initialization for ASR encoders.
\item \textbf{Benefits of Fine-Tuning}: Qwen models that undergo optimization with the ASR loss perform better compared to their frozen counterparts. This underscores the importance of task-specific fine-tuning to further unlock the potential of pre-trained models and enhance their performance in ASR applications.
\end{enumerate}
Collectively, these results not only validate the efficacy and adaptability of transformers from pre-trained LMs in ASR tasks but also emphasize the critical role of fine-tuning these models to optimize their performance for the specific task.

Delving deeper into the influence of pre-trained language model (LM) scale, we introduce a larger variant, Qwen-1.5-1.8B~\footnote{https://huggingface.co/Qwen/Qwen1.5-1.8B}, as the ASR encoder. We replicate Experiments 2 and 3 from Table~\ref{tab1} with this larger model to explore its impact. The outcomes, reflected in Experiments 6 and 7 of Table~\ref{tab1}, reveal that escalating the size of the pre-trained LM does not necessarily lead to enhanced ASR performance.  Conversely, we observe performance improvements when a higher dropout rate (set at 0.5) is applied within the transformer layers. This adjustment strategy, which promotes regularization and mitigates overfitting, points to the possibility that the pre-trained LM might be overly complex or prone to overfitting for the relatively smaller and potentially less diverse AISHELL-1 dataset. These results advocate for a nuanced consideration of model size, dropout usage, and the interplay between pre-training and task-specific fine-tuning when integrating pre-trained LMs into ASR systems.
\begin{table}
\centering
\caption{CER Results on AISLELL-1.}\label{tab1}
\begin{tabular}{l|l|c|c |c}
\hline
\multirow{2}{*}{\textbf{Exp ID}} & \multirow{2}{*}{\textbf{ASR Encoder}} & \textbf{Freeze} & \textbf{Trainable/Total} &  \multirow{2}{*}{\textbf{Dev/Test}}\\
 & & \textbf{Qwen} & \textbf{Parameters} &  \\
\hline
1 & Conv input layer & - & 10M / 10M & 61.95 / 66.21 \\
\hline
2 & + random initialized Qwen  & \Checkmark &	 10M / 318M  & 33.67 / 37.24 \\
3 & + pre-trained Qwen &  \Checkmark	& 10M / 318M & 21.52 / 23.71 \\
\hline
4 & + random initialized Qwen  &  \XSolidBrush & 318M / 318M & 14.05 / 15.56 \\
5 & + pre-trained Qwen   & \XSolidBrush & 318M / 318M & 10.19 / 11.09 \\
\hline
6 & 2 with Qwen1.5-1.8B &   \Checkmark & 19M / 1.23B & 25.27 / 28.15 \\
7 & 3 with Qwen1.5-1.8B &  \Checkmark & 19M / 1.23B & 21.33 / 24.03 \\
\hline
8 & 4 with dropout &  \XSolidBrush & 318M / 318M & 13.47 / 14.54 \\
9 & 5 with dropout & \XSolidBrush & 318M / 318M & 6.89 / 7.71 \\
\hline
\end{tabular}
\end{table}

\begin{table}
\centering
\caption{WER Results on Librispeech.}\label{tab2}
\begin{tabular}{l|l|c|c|c}
\hline
\multirow{2}{*}{\textbf{Exp ID}} & \multirow{2}{*}{\textbf{ASR Encoder}} & \textbf{Freeze} & \textbf{Trainable/Total} &  \textbf{Test}\\
 & &\textbf{Qwen} & \textbf{Parameters} & \textbf{clean/other} \\
\hline
1 & Conv input layer & - & 11M / 11M & 109.97 / 109.94 \\
\hline
2 & + random initialized Qwen  & \Checkmark &	 11M / 319M  & 71.38 / 78.73 \\
3 & + pre-trained Qwen &  \Checkmark	& 11M / 319M & 49.64 / 59.96 \\
\hline
4 & + random initialized Qwen  &  \XSolidBrush & 319M / 319M & 15.65 / 27.94 \\
5 &+ pre-trained Qwen   & \XSolidBrush & 319M / 319M & 3.90 / 8.77 \\
\hline
\end{tabular}
\end{table}

Extending our exploration, we examine if transformers from pre-trained LMs can augment an already proficient ASR encoder's capabilities. Our investigation encompasses three configurations:

(1) Stand-Alone Pre-Trained SAN-M Encoder: This serves as our baseline, utilizing the pre-trained SAN-M encoder without additional Qwen transformer layers.

(2) Frozen Pre-Trained Qwen on SAN-M: Here, we stack frozen pre-trained Qwen transformers on top of the SAN-M encoder.

(3) Unfrozen Pre-Trained Qwen on SAN-M: In this setup, the pre-trained Qwen transformers are integrated with the SAN-M encoder and are further fine-tuned during training.

All models undergo training on our extensive 60,000-hour Mandarin dataset and are evaluated across two test sets: a general test suite comprising diverse speech domains and a specialized test set featuring ancient Chinese poetry, which is based on recitations of poems from classical Chinese literature that often present unique challenges due to their reliance on archaic vocabulary and phrasing, making recognition based solely on pronunciation difficult.

The outcomes detailed in Table~\ref{tab3} reveal that the inclusion of Qwen transformers yields noticeable improvements. Specifically, the frozen Qwen layer brings about a 2\% reduction in Character Error Rate (CER) on the general test set, while the unfrozen version further enhances this to a 7\% decrease. More prominently, on the ancient Chinese poetry test set, the frozen Qwen configuration results in an 18\% CER reduction, with the unfrozen version achieving an impressive 47\% decrease. These substantial gains may be attributed to the enhanced semantic modeling capabilities provided by Qwen, which are particularly crucial for accurately transcribing the context-heavy language of ancient poetry.

\begin{figure}
\centering
\includegraphics[width=0.95\textwidth]{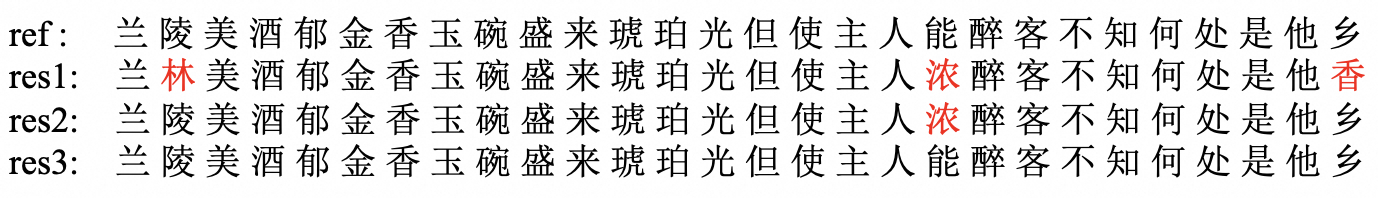}
\caption{Predictions of different models. ref: the reference. res1: Stand-Alone pre-Trained SAN-M encoder. res2: frozen pre-trained Qwen on SAN-M. res3: Unfrozen pre-Trained Qwen on SAN-M.} \label{comp}
\end{figure}

For illustrative purposes, Figure~\ref{comp} showcases sample transcriptions generated by each model, highlighting the differences in prediction accuracy and the clear advantage of integrating Qwen's transformers for complex linguistic tasks.
\begin{table}
\centering
\caption{CER Results on In-house 60000-hour dataset.}\label{tab3}
\begin{tabular}{l|c|c |c|c}
\hline
\multirow{2}{*}{\textbf{ASR Encoder}} & \textbf{Freeze} & \textbf{Trainable/Total} &  \textbf{General} & \textbf{Ancient Chinese}\\
& \textbf{Qwen} & \textbf{Parameters} & \textbf{test} & \textbf{poetry test}\\
\hline
pre-trained SAN-M & - & 160M / 160M & 8.98 & 8.40 \\
\hline
+ pre-trained Qwen &  \Checkmark	& 160M / 470M & 8.80 & 6.92 \\
+ pre-trained Qwen  & \XSolidBrush & 470M / 470M & 8.35 & 4.42 \\
\hline
\end{tabular}
\end{table}
\section{Conclusion and future work}
This research offers early insights into the application of transformers from pre-trained language models (PLMs) within Automatic Speech Recognition (ASR), presenting empirical evidence that highlights their potential as universal representation encoders, transcending their original text-based training to enhance understanding in non-text domains. Especially, our findings underscore the promise of these transformers in tasks demanding strong semantic comprehension, indicating their adaptability and versatility beyond their initial purpose. 
However, our study also identifies two main areas for future exploration:

\textbf{Theoretical Understanding}: While we observe empirical benefits, the underlying theoretical rationale behind why transformers, trained solely on text, can contribute positively to acoustic modeling remains elusive. This poses an intriguing challenge for future research to delve deeper into the intrinsic properties of transformers pre-trained on massive datasets, unraveling how they manage to generalize across modalities from text to speech.

\textbf{Efficiency and Scalability}: The addition of pre-trained LM transformers onto a pre-trained ASR encoder, while showing improvements, exhibits only marginal gains on standard test sets. Considering the substantial increase in model size with limited returns, future endeavors should aim at optimizing the integration strategy. Researchers must explore methods to leverage the power of PLMs more efficiently, striking a balance between model complexity, performance improvement, and computational cost.

\end{document}